\documentclass{article}
\usepackage{amsfonts,amssymb,hyperref}
\usepackage{times,mathptm,anysize}
\usepackage{amsmath,amsthm,graphicx,color}

\begin{document}
   \newtheorem{theorem}{Theorem}
   \newtheorem{lemma}{Lemma}
   \newtheorem{definition}{Definition}
   \newtheorem{property}{Property}
   
   \theoremstyle{remark}
   \renewenvironment{proof}{\noindent{\it Proof.}\;}{$\blacksquare$}
   \newtheorem{example}{Example}

\newcommand{\gc}[1]{\textcolor{green}{\bf #1}}
\newcommand{\rc}[1]{\textcolor{red}{\bf #1}}
\marginsize{1in}{1in}{1in}{1in}

\title{Techniques for the Synthesis of Reversible Toffoli Networks}

\author{
{\small
\begin{tabular}[t]{c@{\extracolsep{4em}}c@{\extracolsep{4em}}c}
D. Maslov &   D. M. Miller            & G. W. Dueck \\
Dept. of Combinatorics and Optimization & Dept. of Computer Science & Faculty of Computer Science \\
University of Waterloo & University of Victoria  & University of New Brunswick \\
Waterloo, ON, Canada & Victoria, BC, Canada & Fredericton, NB, Canada \\
dmitri.maslov@gmail.com & mmiller@uvic.ca       & gdueck@unb.ca \\
\end{tabular}
}
}

\maketitle

\begin{abstract}
This paper presents novel techniques for the synthesis of
reversible networks of Toffoli gates, as well as improvements to
previous methods.  Gate count and technology oriented cost metrics are used.
Our synthesis techniques are independent of the cost metrics.
Two new iterative synthesis procedure employing
Reed-Muller spectra are introduced and shown to complement earlier
synthesis approaches.  The template simplification suggested in
earlier work is enhanced through introduction of a faster and 
more efficient template application algorithm, updated (shorter) classification
of the templates, and presentation of the new templates of sizes 7 and 9.
A novel ``resynthesis'' approach is introduced wherein a sequence
of gates is chosen from a network, and the reversible
specification it realizes is resynthesized as an independent
problem in hopes of reducing the network cost.
Empirical results are presented to show that the methods are effective
both in terms of the realization of all $3\times 3$ reversible
functions and larger reversible benchmark specifications.
\end{abstract}

\section{Introduction}
The synthesis of reversible networks has received much attention
in recent years
\cite{co:tk, co:aj, co:k, ar:mdCAD, ar:mdmCAD, ws:mil, ws:mp, ar:spmh}.
There are two primary motivations for this.  One is power
consumption. Landauer \cite{ar:lan} showed that irreversible
circuits must consume power, and consequently dissipate heat,
whenever they erase or otherwise discard information.  Further,
Bennett \cite{ar:ben} showed that for power not to be dissipated
in an arbitrary circuit, it must be built from reversible gates.
While the heat generation due to the information loss in modern
CMOS is still small, recent work by Zhirnov {\em et al.}
\cite{ar:zkhb} shows the potentially prohibitive difficulty of
heat removal with the increasing density of CMOS. The second
motivator is that all quantum gates are reversible \cite{bk:nc}.

Hence there are compelling reasons to consider circuits composed
of reversible gates and the synthesis of such networks. Reversible
circuit techniques are of direct interest in low-power CMOS design
\cite{phd:schrom}, quantum computing \cite{bk:nc}, and nanotechnology
\cite{ar:m, ar:m93}. Quantum computing seems to be the most promising 
technology in terms of its potential practical use. As a tribute to 
this fact, we wrote our software with an option of minimizing the 
gate count or a quantum cost (in fact, any weighted gate count type 
cost) of the resulting implementation.  
Research on reversible synthesis is of
particular importance to the development of quantum circuit
construction (in particular, oracles) and may well result in much more 
powerful computers and computations.

In this paper, we develop a 
set of techniques for the reversible circuit synthesis and present 
a CAD tool. Due to the small size 
of the modern quantum processor (state of the art quantum processor 
can work with 7 qubits \cite{www:quant}; and, there is a limited 
control over a 12-qubit processor \cite{ar:nmrd}), difficulty in constructing 
a reliable implementation
of the gates in existing hardware, quantum errors and decoherence, this 
is how we addressed the CAD tool designer's challenge:
\begin{enumerate}
\item {\em Reliability:} We present a synthesis approach and its software 
realization that always finds a solution (network). We motivate it such 
that for the people using 
a CAD tool, it is important to get a network no matter how ``difficult''
the function they synthesize is.
\item {\em Scalability:} Our software can be applied to the functions 
with up to 21 variables in reasonable time. While this number is not large, 
it is 3 times (almost twice in case of \cite{ar:nmrd} and limited control) 
greater than the size of the best modern quantum processor. 
This is more than enough for the present needs. 
We store a function as a truth table which has to 
fit in memory --- this limits the scalability of our approach. In 
Section \ref{sec:fw} we indicate how to improve the existing software so as to allow 
synthesis of larger specifications. To date, we did not find it useful to pay much
attention to further scalability.
\item {\em Quality:} Small networks are always in favor, especially on the 
early stage of the development of a technology. Specifics of quantum technology 
include limited computational time due to decoherence and inaccuracy in applying 
the gates leading to accumulation of the errors, among a number of 
other issues. Thus, it is much more important to create
smaller designs for quantum technology, as compared to, for instance, CMOS. Most 
of our attention has been put to decrease the cost of the final implementation. 
Results shown in Section \ref{sec:r} indicate that we succeeded in this direction.
\item {\em Runtime:} Some of our designs may take up to 12 hours to synthesize on an 
Athlon 2400XP machine with 512M of RAM memory running Windows. However, in Section \ref{sec:r}
we discuss how to speed up our tool 6 times on a 6-processor parallel machine.
Optimization of the code (which, in its present form is not optimized),
using a newer compiler (ours is as of 1996), and a more recent computer system would 
also contribute to the runtime reduction. We found that our present realization 
satisfies the market needs as is, in the sense that 12 hours for synthesis compare 
favorably to the 4 years of no progress in the development of larger quantum processors. 
\end{enumerate}

In this paper, we present novel techniques for the synthesis of
reversible networks of Toffoli gates as well as improving on
some existing techniques.  Section \ref{sec:back} provides the
necessary background.  In Section \ref{sec:s1}, we present a new
synthesis approach which selects Toffoli gates so that the
complexity of the Reed-Muller spectra specifying the reversible
function is iteratively reduced until the specification becomes
the identity. The complexity is based on the number of nonzero
coefficients in the spectra.  This method does not always find a
solution, but it frequently finds better solutions than those
found by earlier methods such as the one presented in
\cite{ar:mdmCAD}. We follow this section by description of a
second Reed-Muller spectra based synthesis algorithm (Section
\ref{sec:s2}). A significant advantage of this algorithm is its
guaranteed convergence, and lesser quantum cost in the worst case scenario 
as compared to the previously presented methods \cite{ar:spmh, ar:mdmCAD}. 
Together the new Reed-Muller techniques
and the earlier approach in \cite{ar:mdmCAD} yield significantly
improved results.

As presented in \cite{ar:mdmCAD}, once an initial network is found,
it can often be simplified through the application of templates. In
Section \ref{sec:td}, we present an improved approach to templates
including classification of the templates of size up to 7 and 
some useful templates of size 9. We noticed that the template matching algorithm
of \cite{ar:mdmCAD} is not very efficient, and replace it with a new one.
Our new matching algorithm is better in the sense that, unlike the previous algorithm,
under certain conditions it is guaranteed to find all possible network reductions
that such a templates based tool can find, plus, it works faster.

A new ``resynthesis'' approach is presented in Section
\ref{sec:resynth}. This method depends on the fact that any
sequence of gates in a reversible network on its own realizes a
reversible specification. The method randomly (under some
constraints) selects a sequence of gates from a network and then
applies synthesis methods and the templates to the reversible
function defined by that sequence.  If the network found by
resynthesis is smaller, it replaces the selected sequence in the
original network.  While our current approach to resynthesis is
rather naive, it does significantly reduce the size of the network
in many instances, particularly for some of the larger benchmark
problems.

Empirical results are given in Section \ref{sec:r}.  Our methods
are shown to produce an excellent overall average for the
synthesis of all $3\times 3$ reversible functions, only $0.16\%$
above the optimum. We also present the results of applying our
methods to a number of larger benchmark functions. The paper
concludes with suggestions for ongoing research.

\section{Background}\label{sec:back}

\begin{definition}
An $n$-input, $n$-output, totally-specified Boolean
function $(y_1,y_2,...,y_n)=f(x_1,x_2,...,x_n)$ is {\bf
reversible} if it is a bijection, {\em i.e.} each input pattern is
mapped to a unique output pattern.
\end{definition}

Using methods such as in \cite{ar:mdCAD,ws:mil,co:tk} a (possibly
incompletely-specified) multiple-output Boolean function can be
transformed into a reversible function. These methods are not
particularly efficient and it is an open research problem to
find better ways to perform such a transformation while minimizing
the overhead due to addition of ``constant inputs'' and ``garbage
outputs'' \cite{ar:gc}.  In this work, we assume a reversible
specification as the starting point.

Given a reversible specification, there are many ways ({\em e.g.}
\cite{co:tk,ws:mp,co:aj,co:k,ar:mdmCAD}) of constructing a reversible network
using the multiple control Toffoli gates defined as follows:

\begin{definition}
For the domain variables $\{x_1, x_2, ..., x_n\}$ the {\bf
multiple control Toffoli gate} has the form $TOF(C;t)$, where
$C=\{x_{i_1}, x_{i_2}, ..., x_{i_k}\},\;t=\{x_j\}$ and $C \cap t =
\emptyset$. It maps the Boolean pattern $(x^0_1, x^0_2, ...,
x^0_n)$ to $(x^0_1, x^0_2, ..., x^0_{j-1},$ $x^0_j\oplus
x^0_{i_1}x^0_{i_2}...  x^0_{i_k}, x^0_{j+1}, ..., x^0_n)$.  The
set $C$ which controls the change of the $j$-th bit is called the
set of {\bf controls} and $t$ is called the {\bf target}.
\end{definition}

The most commonly used such gates are: the NOT gate (a multiple control
Toffoli gate with no controls) denoted $TOF(x_j)$, the
CNOT gate (a multiple control Toffoli gate with a single control bit) which
is also known as a Feynman gate \cite{ar:fey} and is denoted
$TOF(x_i;x_j)$, and the original Toffoli gate (a multiple control
Toffoli gate with two controls) denoted $TOF(x_{i_1}, x_{i_2};
x_j)$ \cite{ar:tof}.

A reversible network is composed of reversible gates, which due to
the restrictions dictated by the target technologies \cite{bk:nc}
form a cascade.

\subsection{Cost of a Reversible Toffoli Network}

It is a common practice in reversible logic synthesis 
area \cite{co:tk,ws:mp,co:aj,co:k,ar:mdmCAD} to synthesize a network
using multiple control Toffoli gates and report its cost as a number of gates in it. 
However, from the point of view of technological realization, multiple control
Toffoli gates are not simple transformations. Rather they are
composite gates themselves and Toffoli gates with a large set of
controls can be quite expensive \cite{ar:bbcd, co:mymd}. We point out 
that there are three distinct Toffoli gate simulations \cite{ar:bbcd}, 
one with an exponential cost and requiring 
no auxiliary bits, and two with linear costs and requiring $1$ and 
$n-3$ auxiliary bits for an $n$-bit Toffoli gate. Due to its 
exponential size and usage of infinite number of gate types requiring 
very accurate hardware realization due to the small rotation angles, we 
find multiple control Toffoli gate simulations with zero auxiliary bits impractical. 
Among the remaining two linear cost realizations of the Toffoli gates 
the one associated with availability of $n-3$ auxiliary bits is 
smaller. \cite{co:mymd} improves over the Toffoli gate 
simulation of \cite{ar:bbcd} using the basis of 
elementary quantum gates NOT, CNOT, controlled-$V$ and its inverse 
controlled-$V+$ \cite{bk:nc}. Such quantum gates were studied in the literature
and were efficiently simulated in liquid state NMR (nuclear magnetic
resonance) quantum technology \cite{ar:llklbp}.

\begin{definition}
{\bf Cost} of a Toffoli network with $n$ inputs/outputs is a sum of 
costs of its gates, which may sometimes be followed by an asterisk. 
\begin{itemize}
\item For a network with Toffoli gates of maximal size $n-1$, 
each $k$-bit ($k\leq n-1$) Toffoli gate cost is a minimum of the two linear cost 
realization gate counts \cite{co:mymd} as long as
all associated auxiliary bits can be accommodated in the circuit.
In this case, we do not use asterisk. 
\item For an $n$-bit network containing an $n$-bit Toffoli gate, during
the calculation of the cost of each multiple control Toffoli gate we assume 
presence of an additional auxiliary bit (that is, assume that the 
network is built on $n+1$ wires). In such case, numeric value of the 
network cost is followed by an asterisk.
\end{itemize}
\end{definition}

The lesser numeric value of $cost*$ means a better realization.

In this paper, we report two sets of the synthesis results. In one,
we minimize the gate count. This is done to compare the 
quality of our new approach to the quality of the previously presented 
methods. The second set of results contains networks synthesized 
as to minimize the quantum cost defined above. 
In our software implementation, costs of the multiple control Toffoli 
gates are stored in a table. This allows an easy change 
of the cost we use to direct the circuit simplification into any other 
linear cost network metric.  To our knowledge, this is the first 
attempt in the area of reversible logic synthesis to minimize a technological
implementation cost instead of the gate count. We believe that 
network realizations from the second set are more practical.

\subsection{Reed-Muller Spectra}

Every Boolean function $y=f(x_1,x_2,...,x_n)$ can be uniquely
written as a polynomial of the form $a_0 \oplus a_1x_1 \oplus
a_2x_2 \oplus a_3x_1x_2 \oplus ... \oplus a_{2^n-1}x_1x_2...x_n$
with Boolean coefficients $a_0,$ $a_1, ...,a_{2^n-1}$, which is
referred to as the ``positive polarity Reed-Muller expansion.'' A
compact way to represent this expression is the vector
(``spectrum'') of coefficients $(a_0, a_1, ...,a_{2^n-1})$. Given
a size $n$ reversible function, its Reed-Muller spectra ({\bf RM spectra})
can be viewed as a table of size $n\times 2^n$, where each column
represents the Reed-Muller spectrum of the corresponding output of
the reversible function. Note, that for reversible functions the
last row of this table is all zeroes and the size of the table can
be reduced to $n\times (2^n-1)$.  RM spectra can be
efficiently computed using fast transform techniques similar to a
discrete FFT. The
transformation can be expressed in matrix form \cite{bk:tdm} as

\begin{eqnarray*}
  R &=& M^{n} F\\
  M^{0} &=& \left[ 1 \right] \\
  M^{n} &=& \left[%
\begin{array}{cc}
  M^{n-1} & 0 \\
  M^{n-1} & M^{n-1} \\
\end{array}%
\right]
\end{eqnarray*}

\noindent where the summation is modulo-2, {\em i.e.} EXOR, 
and $F$ is the truth vector of the given function. 
In our software, this transformation is implemented by the code shown
below which maps a truth vector f[\;] of length $2^n$ given as an
array of integers into the RM spectrum for the given function.

\begin{verbatim}
void RMT(int f[]){
  int i,j,k,m,p;
  int n = log(LengthOfVector(f[]));

  for (m=1;m<(2*n);m=2*m)
    for (i=0;i<2^n;i=i+2*m)
      for (j=i,p=k=i+m;j<p;j=j+1,k=k+1)
        f[k] = f[k] ^ f[j]; // bitwise EXOR
}
\end{verbatim}
The elements of f[\;] can be multi-bit values with each position
representing a separate output function, in which case the
procedure computes the output function RM spectra in parallel.
Computation of RM spectra in this way is quite efficient for
problems with a number of outputs up to the number of bits in an
integer for the computer and compiler used.

Important properties of this transformation include:
\begin{enumerate}\label{RMTp}
    \item {\em self
inverse} {\em i.e.} $RMT(RMT(f))=f$;
    \item {\em order dependence} in the
sense that value $f[k]$ is never updated using a value $f[j]$
where $j \geq k$;
    \item {\em power-of-two independence}
in the sense that value $f[k]$ for $k=2^s$ is never updated with
values of $f[j]$, where $j=2^t$ and $1 \leq s,t \leq n$.
\end{enumerate}

The RM spectra of the size $n$ identity function with
outputs $y_1,y_2,...,y_n$ has a single nonzero coefficient
$a_{2^{i-1}}$ for each $y_i$ with all other coefficients 0.

\begin{definition}
The {\bf RM cost} of a reversible function is the total number of
coefficients for which its spectra differs from the spectra of the
identity function.
\end{definition}

We will refer to each nonzero row of the tabular representation of
the RM spectra for the identity function as a {\bf variable row}.
Such variable rows are  those at positions $1,2,...,2^{n-1}$. We
will also refer to all others as {\bf non-variable rows}.

\subsection{Direct Application of a Toffoli Gate in RM Spectra}

Application of a multiple control Toffoli gate $TOF(x_{i_1}, x_{i_2},
...,x_{i_k};x_j)$ from the input side of a reversible
specification simply requires replacing each occurrence of the
literal $x_j$ in the Reed-Muller expansion of the output variable
$y_s= a_{0} \oplus a_{s,1}x_1 \oplus a_{s,2}x_2 \oplus
a_{s,3}x_1x_2 \oplus ... \oplus a_{s,2^n-1}x_1x_2...x_n$ with the
expression $x_j \oplus x_{i_1}x_{i_2}... x_{i_k}$ followed by
simplification of the resulting expression. In the case where the
Reed-Muller spectra is stored as a table this operation requires
at most $n\times 2^n$ binary operations with no algebraic
simplification. Application of a multiple control Toffoli gate
$TOF(x_{i_1}, x_{i_2}, ...,x_{i_k};x_j)$ from the output side can
also be done directly in the spectra. In particular, the
polynomials given by columns $y_{i_1}, y_{i_2},...,y_{i_k}$
(Boolean vectors of length $2^n$) of the RM spectra are multiplied
and the result is EXORed with column $y_j$ with the result
stored in column $y_j$. Hence, a Toffoli gate can be 
applied directly while working with the RM spectra. We note that since most
reversible functions have numerous zero rows in their tabular RM
spectra, it may be more efficient to store the indices and
values of non-zero elements of the RM spectra. In this case,
application of a Toffoli gate may require significantly less
operations.  This needs to be pursued but to date we have found
the tabular approach sufficient for our work.

\section{Iterative Network Synthesis Using Reed-Muller Spectra} \label{sec:s1}

The first synthesis algorithm that we propose is very simple. At
each step, by exhaustive enumeration it selects the Toffoli gate
whose application to the function specification results in the 
largest decrease of the RM cost. If no gate application
decreases the RM cost, a gate is chosen that
results in the minimal increase of the RM cost. In both cases, if
there is a tie between two or more gates, a gate with the smallest
control set is chosen. If there is a tie based on number of
controls, our method selects the first gate in lexicographic
order.

This synthesis approach is similar to some of earlier proposed
techniques \cite{ws:mil, ar:mdCAD} in that the gates are assigned 
to decrease some sort of function complexity measure. However,  
we use a different gate library and here
the Reed-Muller spectrum is used rather than the Walsh spectrum \cite{ws:mil}
or Hamming distance defined over the truth table \cite{ar:mdCAD},
resulting in significantly better synthesis results. This is
because the Reed-Muller spectrum better corresponds to the
functional operation of Toffoli gates.

It is not surprising that there are drawbacks to such a simple
approach.  When considering larger benchmark specifications, we
identified two major problems. First, the algorithm is not
guaranteed to converge.  In particular, among the functions we tried, 
it did not converge for the $hwb$ type benchmark functions with 
more than 5 variables and function $ham7$. We address this problem by
using the other algorithm (that always converges)
first and taking its gate count as the upper bound for the
synthesis using this algorithm. If the algorithm based on minimization 
of RM cost does not return a network with fewer gates than
the second algorithm that we present below, its solution is considered
inefficient. While such a technique appears efficient for the
synthesis of benchmark functions, we are working on heuristics to
force the Reed-Muller based method to converge on every reversible
specification.

The second major problem with the new algorithm is that at each
step it tries every possible Toffoli gate, of which there are
$n\times 2^{n-1}$ for a size $n$ reversible function. Current
implementation of this algorithm uses a table to store the values
of RM spectra, making the cost of the search for a single Toffoli
gate assignment equal $n^2\times 4^{n-1}$ binary operations. In
practice, it is likely too time consuming to synthesize functions
with more than 12 input variables (especially if the resynthesis
technique discussed below is also used). We addressed this issue 
by having an option to limit the number of controls which every 
gate that we try might have.
We plan to improve the runtime further 
by first exploring the idea noted above of storing only the
nonzero coefficients; and second, searching for heuristics that
can guide the selection of a Toffoli gate to avoid the current
exhaustive enumeration. Ideas presented in \cite{co:aj,co:k} 
might be useful. Further, we developed another synthesis
algorithm that does not have these two major problems. This
algorithm is outlined in the next section.

Despite the above deficiencies, the new iterative algorithm by itself converged 
for every one of the 40,320 $3\times 3$ reversible functions. 
It synthesized them with an average of 6.38
Toffoli gates per function in 3 seconds on a 
single 750 MHz processor Sun Blade 1000. This compares
quite favorably to the 7.25 average \cite{ar:mdmCAD} for the MMD
algorithm with no templates applied and shows that the new method
has very good potential.

\section{MMD Type Reed-Muller Spectra Based Synthesis Approach} \label{sec:s2}

The second Reed-Muller based algorithm that we present
is similar to MMD \cite{ar:mdmCAD} in the sense that it works with a
single row at a time, and allows a similar bidirectional
modification. However, there are a number of differences between
MMD and the new algorithm. Some of them are:
\begin{itemize}
\item our new algorithm works with Reed-Muller spectra, not in the
Boolean domain (truth table) as does MMD; 
\item the choice of gates while working with a
single row is completely different; 
\item at any point MMD does
not change the correct form of upper rows, which is not true for
the new method.
\end{itemize}

We start with a description of the unidirectional (basic) version
of the algorithm. It consists of $2^n-1$ steps (numbered $0$ to
$2^n-2$). At each step $i$, the first $i$ rows (rows with numbers
$0,1,...,i-1$) in the tabular RM spectra of the function under
synthesis match the first $i$ rows of the RM spectra of the
identity function. The algorithm assigns a (possibly empty) set of
Toffoli gates such that the $i^{th}$ row of the tabular RM spectra
is transformed to match the $i^{th}$ row of the RM spectra of the
identity function. It can be observed that for such an algorithm,
when step $2^n-2$ is completed, the RM spectra is transformed to
the RM spectra of the identity function, and thus the target
specification is successfully synthesized. This is because the
$(2^n-1)^{st}$ row of the RM spectra of a reversible function is always
zero. We now describe which gates are assigned depending on the
value of $i$ and outline a proof showing that a suitable set of
gates can always be found.  We use $(r_n,r_{n-1},...,r_1)$ to
denote the values in a row of the tabular representation for the
RM spectra for the reversible function under consideration.  We
refer to a row as being {\em earlier} than another if it has a
lower row index number.

{\bf A: Step $i=0$.} This step is unique since it is only for this
step that we use NOT gates, and there is no need to consider if
earlier rows are changed since there are none. Given the 0-th row
has values $(r_n,r_{n-1},...,r_1)$ we apply NOT gates $TOF(x_j)$
for every $r_j=1, \; j=1..n$.

{\bf B: Step $i=2^{k-1},\;k=1..n$.} Each of these rows is a variable row.
Such a row, $(r_n,r_{n-1},...,r_1)$, has to be brought to the form
$(0,0,...,0,1,0,...,0)$ with 1 at a position $k$. This is done
through the following two procedures. We first check if $r_k=1$.
If it is not, we make it equal one by assigning a gate
$TOF(x_s;x_k)$ such that $s=\max\{j|\; r_j=1, \; j=1..n\}$ and
$s>k$. According to Lemma \ref{lem:uni} such an $s$ exists, and it
can be easily verified that application of the gate $TOF(x_s;x_k)$
does not affect RM spectra rows earlier in the table.

At this step the row we are working with has the form
$(r_n,r_{n-1},...,r_{k+1},1,r_{k-1},...,r_1)$. We next use gates
$TOF(x_k;x_j)$ for every $r_j=1, \; j=1..n$. By applying such
gates we do not change rows earlier in the table than the row we
are working with and at the same time the $i^{th}$ row is
transformed to the desired form $(0,0,...,0,1,0,...,0)$ with 1 at
position $k$.

{\bf C: Step $i,\; i \neq 2^k,\; i>0$.} For these $i$, we know
that we are working with a non-variable row. Assume it has
values $(r_n,r_{n-1},...,r_1)$. It has to be transformed to the
form $(0,0,...,0)$, which is the form of each non-variable row of
the RM spectra of the identity function. We first find
$s=\max\{j|\; r_j=1, \; j=1..n\}$ and such that item $2^{s-1}$ does
not appear in the binary expansion of $i$. In other words, choose
a variable whose $i^{th}$ value in the RM spectra is 1 and that is
not included in the product associated with the $i^{th}$ element
of the RM spectra. Such an $s$ exists according to Lemma
\ref{lem:uni}. We first apply gates $TOF(x_s;x_j)$ for every
$r_j=1, \; j \neq i, \; j=1..n$. This transforms the row we are
working with into $(0,0,...,0,1,0,...,0)$ with 1 at position $s$.
Second, we apply gate $TOF(X_i;x_s)$, where $X_i$ is a product of
variables $x_j$ such that the $j^{th}$ bit of the binary expansion
of the number $i$ equals one. Such an operation transforms the row
we are working with into the desired $(0,0,...,0)$. Finally, we
undo $TOF(x_s;x_j)$ if such gates changed RM spectra rows earlier
in the table than row $i$. Clearly, such ``undo'' operations do
not change the correct form of the pattern we are working with.

To complete the proof of convergence for the above algorithm we
need to show that at steps B and C a value $s$ with the required
properties can always be found. The following Lemma proves this.

\begin{lemma} \label{lem:uni}
Suppose the RM spectra of a reversible function $f$ has its first
$i$ rows (rows with numbers $0,1,...,i-1$) equal to the first $i$
rows of the identity function. Denote the $i^{th}$ row value by
$(r_n,r_{n-1},...,r_1)$. Then,
\begin{itemize}
\item If $i=2^{k-1}$ $(k=1..n),$ then $(r_n,r_{n-1},...,r_1) \neq
(0,0,...,0)$. 
\item If $i=2^{k-1}$ $(k=1..n)$ and $r_k=0$, then
the number $s$ defined as $\max\{j|\; r_j=1, \; j=1..n\}$ is
greater than $k$. 
\item If $i\neq 2^k$ and $(r_n,r_{n-1},...,r_1)
\neq (0,0,...,0)$, then there exists $s$, $1 \leq s \leq n$ such
that $2^{s-1}$ does not appear in the binary expansion of $i$ and
$r_s=1$.
\end{itemize}
\end{lemma}

\begin{proof}
First, we prove by contradiction that if $i=2^{k-1}$, then
$(r_n,r_{n-1},...,r_1) \neq (0,0,...,0)$. Suppose
$(r_n,r_{n-1},...,r_1) = (0,0,...,0)$ and apply RMT. RMT will
transform $(r_n,r_{n-1},...,r_1) = (0,0,...,0)$ at position
$i=2^{k-1}$ into itself due to the properties 2 and 3 (order dependence and 
power-of-2 independence) of the RMT and the fact all
non-variable rows earlier than the $i^{th}$ row are zero.
According to the property 1 (self inverse) of RMT we are in the
Boolean domain now, and we have two rows, the $0^{th}$ and
$i^{th}$, both equal to $(0,0,...,0)$. This is a contradiction
since a reversible function can not have two equal rows in its
truth table representation.

Proof of the second statement is similarly shown by contradiction
by assuming that such an $s$ (which does exist as a result of the
proof of the first statement) is less than $k$. In that case
$(r_n,r_{n-1},...,r_1)$ can be interpreted as a binary expansion
of a number $C<2^{k-1}$ since its largest digit is at a position
right of $k$. After applying RMT we move to the Boolean domain and
find that pattern $(r_n,r_{n-1},...,r_1)$ did not change. At the
same time, higher in the table, at position $C$, we will find a
pattern equal to $(r_n,r_{n-1},...,r_1)$. This is the
contradiction.

The proof of statement 3 is similar to the above two proofs.
Assume such an $s$ does not exist. Then, $(r_n,r_{n-1},...,r_1)$
may contain ones only at those positions where the binary
expansion of $i=(i_n,i_{n-1},...,i_1)$ has ones. RMT transforms
$(r_n,r_{n-1},...,r_1)$ into $(r_n \oplus i_n,r_{n-1} \oplus
i_{n-1},...,r_1 \oplus i_1)$, a pattern that may have ones only at
positions where the binary expansion of $i$ has ones. An equal
pattern may be found in the truth table at position $i-r$, where
$r$ is an integer with binary expansion $(r_n,r_{n-1},...,r_1)$.
Again having two equal patterns in the truth table is a
contradiction.
\end{proof}

\begin{example}
Consider the 3-variable reversible function specified by the
permutation $[1,0,3,2,5,7,4,6]$ in Boolean domain.  The spectra for this function
are shown in tabular form in the column labelled RM in Table
\ref{tab:ex}. We want to select Toffoli gates to transform this
specification into that of the identity (Table \ref{tab:ex},
column Id).

The first row of the function specification does not match the
first row of the RM spectra of the identity function. This can be
fixed by applying the NOT gate $TOF(a)$. Application of this NOT
gate from the output side transforms the specification into the
one shown in Table \ref{tab:ex}, column S1. The first 5 rows in
specification S1 match the first 5 rows of the RM spectra of the
identity. We need to transform the sixth row from 011 to 000.
First, we decrease the number of ones by applying $TOF(b;a)$. This
leads to specification S2. Note that the third row of S2 has also
changed, which means that it has to be updated later. Next,
transform the sixth row of S2 into the desired form 000 by applying
$TOF(a,c;b)$. This results in specification S3. Finally, undo CNOT
gate $TOF(b;a)$, which leads to the identity specification and
thus the network ($TOF(b;a)\;TOF(a,c;b)\;TOF(b;a)\;TOF(a)$) was
constructed.  We again emphasize that the gates have been
identified from the output to the input.

\begin{table}
\begin{center}
\begin{tabular}{|r|c|c|c|c|c|}\hline
                        & Function  & Step1  & Step2  & Step3  & Id \\ \hline 
Spectral coef. of & cba & cba & cba & cba & cba \\ \hline
1   & 001 & 000 & 000 & 000 & 000 \\
a   & 001 & 001 & 001 & 001 & 001 \\
b   & 010 & 010 & 011 & 011 & 010 \\
ab  & 000 & 000 & 000 & 000 & 000 \\
c   & 100 & 100 & 100 & 100 & 100 \\
ac  & 011 & 011 & 010 & 000 & 000 \\
bc  & 011 & 011 & 010 & 000 & 000 \\
abc & 000 & 000 & 000 & 000 & 000 \\ \hline 
Gate applied: & $TOF(a)\nearrow$ & $TOF(b;a)\nearrow$ & 
$TOF(a,c;b)\nearrow$ & $TOF(b;a)\nearrow$ & \\ \hline
\end{tabular}
\caption{Synthesis of an example function stored as a RM spectra. The result of application 
of the gate on the bottom of each column is shown in the following column while reading from
left to right.}
\label{tab:ex}
\end{center}
\end{table}

\end{example}

\subsection{Bidirectional Method}
The following Lemma suggests how a bidirectional modification can
be developed.

\begin{lemma} \label{lem:bi}
Suppose the RM spectra of a reversible function $f$ has its first
$i$ rows equal to the first $i$ rows of the identity function.
Then, so does the RM spectra of $f^{-1}$, the inverse of $f$.
\end{lemma}
\begin{proof}
This statement is, obviously, correct if one works with the truth
table representation. In particular, if function $f$ maps a
pattern $j$ into itself in the truth table, so will the inverse
function. Due to the property 2 (order dependence) of the RMT
the same holds for all $j,\; 0\leq j<i$.
\end{proof}

Assume the first $i-1$ positions in the RM spectra of $f$ match
the first $i-1$ positions of the RM spectra of the identity, then
according to Lemma \ref{lem:bi} so do the first $i-1$ positions of the RM
spectra of $f^{-1}$. Hence, every assignment of gates that
transforms the $i^{th}$ row of $f$ to match the $i^{th}$ row of
the identity without changing earlier rows (such gates are
assigned from the output side of the cascade) will also transform
the $i^{th}$ row for $f^{-1}$ to match the $i^{th}$ row of the
identity. Analogously, an assignment of gates that ``fixes'' the
$i^{th}$ row for $f^{-1}$ will transform the $i^{th}$ row of the
RM spectra of the identity to its correct form (in this case, the
gates are assigned to the input side of the cascade being built).
The question of which specification to work with, that of the
function, or its inverse is equivalent to the question of which
side of the network to assign the gates to, the input side or the
output side. This is why we call this modification bidirectional.

In our approach, the decision is based on how small is the cost
associated with fixing the $i^{th}$ row of either RM spectra, that
using the function or its inverse. By choosing a smaller cost
associated with such transformation we hope to synthesize an
overall cheaper network. In the case of a tie, we base our
decision on the RM cost of the remaining specification --
preference is given to a set of transformations that yield lower
RM cost. We base this decision on the belief that on average
functions with smaller RM cost are simpler to synthesize. Finally,
when these criteria do not resolve the choice, the gates are
assigned to the output side (working with RM spectra of the
function). Perhaps, better heuristics for the decision of which
side to work with can be found, and it would be both interesting
and beneficial to explore that.

\begin{theorem}
For any reversible function of size $n$ the network synthesized by
either of the two methods (unidirectional or bidirectional)
contains
\begin{enumerate}
\item {\bf In the multiple control Toffoli gates library:} at most $n$ NOT
gates, at most $2(n-1)(2^n-n-2)+n^2$ CNOT gates, and at most
${n\choose k}$ Toffoli gates with $k$ controls for each $k\in
[2..n-1]$. 
\item {\bf In NCT \cite{ar:spmh} library:} at most $n$
NOT gates, at most $2n2^n+o(n2^n)$ CNOT gates, and at most
$3n2^n+o(n2^n)$ Toffoli gates (assuming an additional auxiliary
bit is available; otherwise the circuit may not be constructible
\cite{ar:spmh}). 
\item {\bf In NCV \cite{co:mymd} library:} at most
$11n2^n+o(n2^n)$ NOT, CNOT, controlled-$V$ and controlled-$V^+$
gates (again, assuming an additional auxiliary bit is available;
otherwise the circuit may not be constructible).
\end{enumerate}
\end{theorem}

\begin{proof}
Proof of the first statement is based on an analysis of the basic 
synthesis algorithm described above. First, at most $n$ NOT gates
are used at step 0 ({\bf A:}) of the synthesis algorithm, and none are used
thereafter. At most $n$ CNOT gates are required at each of the
steps $i=2^k$ ({\bf B:}), totalling $n^2$ CNOT gates. At each step $i,\; i
\neq 2^k,\; i>0$ ({\bf C:}) at most $2(n-1)$ CNOT gates are required. The
number of such steps is $2^n-n-2$, giving a grand total of
$2(n-1)(2^n-n-2)+n^2$ CNOT gates. Finally, on each step $i,\; i
\neq 2^k,\; i>0$ ({\bf C:}), assuming $i=2^{i_1}+2^{i_2}+...+2^{i_k}$, at
most one multiple control Toffoli gate with control set
$\{x_{i_1},x_{i_2},...,x_{i_k}\}$ is used. Calculating the number
of such Toffoli gates with $k$ controls gives ${n\choose k}$.

Calculation of the result in NCT library is based on multiple
control Toffoli gate realizations from \cite{ar:bbcd}.  In
NCV library, the result is based on multiple control Toffoli gate
realizations from \cite{co:mymd} and formulas
$$\sum_{k=0}^{n}k{n \choose k}=n2^{n-1};\;\;\;
\sum_{k=0}^{n/2}k{n \choose k}=n2^{n-2}+o(n2^n).$$
\end{proof}

Item 2 of the above theorem shows a lower upper bound (under the
natural assumption that a CNOT gate is no more expensive than a
larger Toffoli gate) for our synthesis algorithm as compared to
the upper bound of $n$ NOT gates, $n^2$ CNOT gates and
$9n2^n+o(n2^n)$ Toffoli gates for the synthesis algorithm in
\cite{ar:spmh}. We also note another feature of this algorithm that
might be useful for a more robust algorithm implementation. 
Linear reversible functions will always be synthesized using NOT 
and CNOT (linear) gates only. While synthesizing linear functions,
it is sufficient to store only zeroth and all variable rows of 
its RM spectra. This allows synthesis of size $1000\times 1000$
linear reversible functions while making a minimal change to the 
existing software.

The synthesis algorithm described in this section 
synthesized all size 3 reversible functions with 
the average of 6.98 Toffoli gates per function (no templates applied)
in 7 seconds on AMD Athlon 2400+ processor. 
Again, this compares favorably to the 7.25 average for MMD algorithm.
For larger benchmark specifications this synthesis 
algorithm usually constructs smaller quantum cost network as 
compared to MMD and iterative algorithm discussed in the previous section.

\section{Comparison of the New Methods with MMD}
In this section we compare performance of the newly presented synthesis methods 
to the performance of MMD method \cite{ar:mdmCAD}. Table \ref{pt} lists the 
name and size of a benchmark function tested and the number of gates and 
quantum cost calculated when synthesis methods MMD, new iterative RM spectra based 
and new MMD-type RM spectra based are applied. Based on this 
test we make the following conclusions. The iterative RM spectra based method 
generally produces smallest circuits for small specifications. However, when tested 
on larger functions it may diverge (Div.) or take a long time to 
complete, and thus does not apply (N/A). For larger specifications, 
RM spectra based MMD type method takes the lead as far as quantum costs are concerned,
and application of the original MMD method results in the smallest gate count. 
In scope of this paper, a smaller quantum cost is more desirable than a smaller gate 
count, because quantum cost is a better indication of the technological cost of 
constructing the circuit.

\begin{table}
\begin{center}
\begin{tabular}{|c|c||c|c||c|c|c|c|} \hline
{\bf name} & {\bf size}& {\bf MMD GC} & {\bf MMD QC} & {\bf Iter. GC} 
& {\bf Iter. QC} & {\bf RM-based GC} & {\bf RM-based QC} \\ \hline
3\_17        & 3  & {\bf 6} & {\bf 14} & {\bf 6}   & {\bf 14}   & 7   & 15 \\ \hline
4\_49        & 4  & 16  & 72* & {\bf 15}  & {\bf 71*}   & 20  & 72* \\ \hline
4mod5        & 5  & 9   & 25      & {\bf 7}   & {\bf 15}   & 9   & 25 \\ \hline
5mod5        & 6  & 18  & 177* & {\bf 12} & {\bf 85*} & 18 & 177* \\ \hline
add3         & 4  & 6   & 18 & {\bf 5} & {\bf 13} & 6  & 14 \\ \hline
cycle10\_2   & 12 & {\bf 19}  & {\bf 1206}  & 27 & 1569 & {\bf 19} & {\bf 1206} \\ \hline
cycle17\_3   & 20 & {\bf 48}  & {\bf 6069}  & N/A & N/A & {\bf 48} & {\bf 6069} \\ \hline
ham3         & 3  & {\bf 6} & {\bf 10}  & 7   & 11   & {\bf 6} & {\bf 10} \\ \hline
ham7         & 7  & {\bf 25} & 93  & Div.  & Div.  & 31 & {\bf 57} \\ \hline
ham15        & 15 & {\bf 138} & 2145    & N/A  & N/A & 159 & {\bf 264} \\ \hline
hwb4         & 4  & 18  & 70*     & {\bf 12}  & {\bf 48*}  & 16 & 56* \\\hline
hwb5         & 5  & 57  & 481*    & 55  & 569* & {\bf 53} & {\bf 183}\\ \hline
hwb6         & 6  & {\bf 134} & 1723*   & Div.  & Div. & 149 & {\bf 816*} \\ \hline
hwb7         & 7  & {\bf 302} & 5528*   & Div. & Div. & 435 & {\bf 3036*} \\ \hline
hwb8         & 8  & {\bf 688} & 15527*  & Div. & Div. & 1101 & {\bf 7699*} \\ \hline
hwb9         & 9  & {\bf 1625} & 48384* & Div. & Div. & 2787 & {\bf 22284*} \\ \hline
hwb10        & 10 & {\bf 3694} & 124022* & Div. & Div. & 6291 & {\bf 49303*} \\ \hline
hwb11        & 11 & {\bf 8312} & 343654* & Div. & Div. & 14566 & {\bf 126709*}\\ \hline
mod5adder    & 6  & 37  & 591* & {\bf 24} & {\bf 242} & 63 & 524* \\ \hline
mod1024adder & 20 & {\bf 55}  & {\bf 1575}  & N/A & N/A & {\bf 55} & {\bf 1575} \\ \hline
rd53         & 7  & 20  & 181   & {\bf 19} & {\bf 113} & {\bf 19} & 181 \\ \hline
\end{tabular}
\caption{Testing performance of the synthesis methods.}
\label{pt}
\end{center}
\end{table}

\section{Templates}\label{sec:td}
In previous sections we discussed how to obtain a Toffoli circuit given a function 
specification. Since optimal synthesis is not feasible, we employed a number of 
heuristics. The result of heuristic search is, usually, a non-optimal circuit 
specification. Thus, optimization techniques can be applied to such circuits. 
In particular, we investigate a form of local 
optimization technique, called the templates.

Templates are a generalization of rewriting rules.
A {\bf rewriting rule} is defined as a procedure that
takes 2 equivalent (computing the same function) circuits
and replaces one with the other. If the cost of the part of the circuit
to be replaced is greater than the cost of the replacement circuit
this leads to a circuit cost reduction. Literature encounters two 
attempts other than the templates suggesting how the rewriting rules
can be used to decrease the gate count in a reversible network 
\cite{ws:iky,ws:sppmh} and one for quantum circuits \cite{quant-ph/0307111}.

In many reversible logic synthesis papers, the cost of a network
is defined as a simple weighted gate count. We refer to this as a
{\bf linear cost circuit metric}. In the more general case of a
{\bf non-linear cost metric}, the cost of the complete circuit
does not relate to the gates in a simple linear manner. An example
of such a situation can arise when considering Peres gate
\cite{ar:peres} which, when simulated in quantum technology by a
Toffoli
gate (cost 5) and a CNOT gate (cost 1) would have a cost of 6, whereas a Peres 
gate constructed directly in terms of quantum primitives has cost 4. When
elementary quantum blocks are decomposed into pulses (as it is
done in NMR quantum technology), similar nonlinear cost effects can arise.

We call a rewriting rule {\bf regular} if the replacement circuit
has smaller cost, otherwise we call it {\bf irregular}. The
qualifier regular is omitted when it is clear in context. The idea
of applying regular rewriting rules to transform sub-circuits of a
given circuit is a powerful tool for circuit cost reduction. 
(Application of irregular rules may be helpful in techniques 
like simulated annealing.) The
simplification procedure consists of two parts. First, find as
many regular rewriting rules as possible, and second, apply them
to reduce the cost of a given circuit. Straightforward application
of such an approach to quantum circuit cost reduction can be found
in \cite{quant-ph/0307111} and was proposed in \cite{ws:iky} for
reversible networks composed of multiple control Toffoli gates. However, there are
potential problems with this approach in its simplest form.
\begin{itemize}
\item The number of regular rewriting rules is
very large even for small parameters. For instance, assuming
Toffoli type gates have unit cost, the number of regular rewriting
rules for reversible binary networks where $k=3$ gates are
replaced with $s=2$ gates in a network with $n=3$ input/output
variables is 180. It can be easily shown that this number grows
exponentially with respect to each of the parameters $k,s$ and
$n$.

\item Often, rewriting rules are redundant in the sense
that a $G_1G_2G_3 \rightarrow G_4G_5$ rewriting rule can be a
derivative of a $G_2G_3 \rightarrow G_5$ rewriting rule if $G_1=G_4$.
Further, it can be shown that even the number of non-redundant rules
grows exponentially on $n$, and, likely grows
exponentially on $k$ and $s$ (keeping $s<k$).

\item It can happen that interchanging the order of the gates in a cascade, 
which is frequently possible and which does not change the linear
term cost of a network, may permit application of a rewriting rule
that decreases the cost.
\end{itemize}

The following observations are useful to understanding template approach.

\noindent {\bf Observation 1.} For any network $G_0G_1...\;G_{m-1}$
realizing function $f$ network $G^{-1}_{m-1}G^{-1}_{m-2}...\;G^{-1}_0$
is a valid network\footnote{Toffoli gates are self inverses: every 
gate $G=G^{-1}$. Thus, template application to the networks with 
Toffoli gates will not require introducing the new gate types.} 
for the function $f^{-1}$. This of course includes the 
case where the cascade of gates realizes the identity in which case the 
inverse function is also the identity.  We use $Id$ to denote both the 
identity function and a network realizing the identity function, the
meaning being clear from the context.

\noindent {\bf Observation 2.} For any rewriting rule
$G_1G_2...\;G_k \rightarrow G_{k+1}G_{k+2}...\;G_{k+s}$, its gates
satisfy the following:
\begin{equation*}
G_1G_2...\;G_kG^{-1}_{k+s}G^{-1}_{k+s-1}...\;G^{-1}_{k+1} = Id.
\end{equation*}

\noindent {\bf Observation 3.} For $G_0G_1...\;G_{m-1} = Id$
and any parameter $p,\; 0 \leq p \leq m$
$G_0G_1...G_{p-1} \rightarrow G^{-1}_{m-1}G^{-1}_{m-2}...\;G^{-1}_p$
is a rewriting rule. In the most {\bf trivial circuit cost metric},
where the cost of every gate is 1, {\em i.e.} the gate
count is calculated, the rewriting rule is regular for parameters
$p$ in the range $\frac{m}{2} < p \leq m.$

\noindent {\bf Observation 4.} If $G_0G_1...\;G_{m-1} = Id$,
then $G_1...\;G_{m-1}G_0 = Id$.

Observation 4 allows one to write
an identity network with $m$ gates in $m$ (generally) different
ways. We refer to each as a {\bf cycle}.
We are now ready to give the formal definition of the templates.

\begin{definition}
A {\bf size $m$ template} is a cascade of $m$ gates (a network)
that realizes the identity function.  For a cascade to be a template, 
there must be at least one cycle of the gates that can not be reduced 
in size (gate count) by application of smaller or equal size templates.  
Only the irreducible cycles are used when applying templates.
A template $G_0\;G_1...\;G_{m-1}$ can be applied in either direction:
\begin{enumerate}
\item {\bf Forward application} is a rewriting rule of the form

$G_iG_{(i+1) \bmod m}...\;G_{(i+p-1) \bmod m} \rightarrow$
$G^{-1}_{(i-1) \bmod m}G^{-1}_{(i-2) \bmod m}...\;G^{-1}_{(i+p) \bmod m}$,
where $0 \leq i,p \leq m-1$.
\item {\bf Backward application} is a rewriting rule of the form \newline
$G^{-1}_iG^{-1}_{(i-1) \bmod m}...\;G^{-1}_{(i-k+1) \bmod m}
\rightarrow
G_{(i+1) \bmod m}G_{(i+2) \bmod
m}...\;G_{(i-k) \bmod m}$, where $0 \leq i,p \leq m-1$.
\end{enumerate}
\end{definition}

\begin{figure}[t]
\centering
\includegraphics[height=40mm]{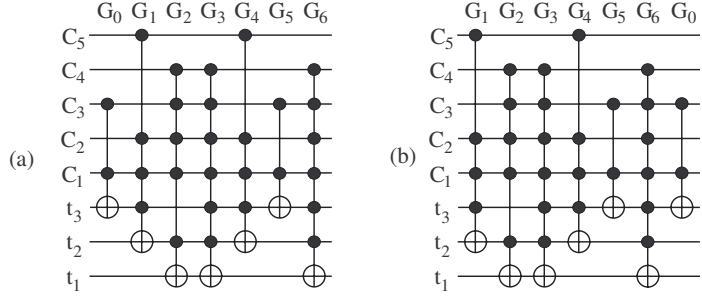}
\caption{Two cycles of a size 7 template: (a) can
not be simplified using smaller and equal size templates and (b)
can be simplified---gates G5, G6, and G0 can be replaced with two
if template (5a) from Figure \ref{alltemplates} is used.} 
\label{tem7inv}
\end{figure}

Our earlier template definitions did not require existence of a
cycle that cannot be simplified, however, this part of the
definition is important. We illustrate this with an example of a
size 7 template with two cycles such that one simplifies and the
other does not, shown in Figure \ref{tem7inv}. 
The network in Figure \ref{tem7inv}(a) does not simplify whereas the 
one in Figure \ref{tem7inv}(b) can be simplified since its rightmost three 
gates can be replaced with two gates $TOF(t_2,t_3,C_1,C_2,C_3,C_4;t1)$ and
$TOF(t_2,C_1,C_2,C_3,C_4;t_1)$.

Correctness of the template definition follows from the above four
observations. One of the immediate benefits that the templates
bring is significant reduction of the space required to store the
rewriting rules (this is a significant improvement considering how much 
space is required in \cite{ws:sppmh} to store some small identities). 
In fact, one template occupies the same storage
space used by a single rewriting rule, yet it is capable of
storing up to $2m^2$ non-redundant rewriting rules. Assuming the
trivial circuit cost metric where each gate has a cost of one, the
number of regular non-redundant rewriting rules can be as high as
$m^2$ for the odd numbers $m$ and $m(m-1)$ for even $m$.

We earlier observed that the number of non-redundant rewriting
rules grows exponentially, therefore template classification is
highly desirable. Depending on the set of model gates,
classifications differ. We consider some of the particular
questions and methods of proper classification of Toffoli
templates in the next section.

\subsection{Toffoli Templates}\label{sec:tc}

We wrote a program that helped us find the Toffoli templates.
To build templates of size $s=s_1+s_2$ the program first uses depth first search
to find optimal networks of maximal sizes $s_1$ and $s_2$ using 3 to 4 input
variables (which likely provides enough generality --- 
we do not have a formal proof that it does --- to find if a template is
missing, but fails to generalize it once a candidate is found). In the
second step, the program computes two sets with the truth vectors of functions
realizable by cascades of sizes $s_1$ and $s_2$. Then, for every
function in the first set it finds its inverse in the second
set. If such a function is found the two networks are
combined (use observations 1 and 2 to see that the resulting cascade
is always the identity function) and templates of size less than $s$
are applied to simplify the cascade. If this leads to a simplification for all cycles,
the constructed identity is not a new template. Otherwise, it is a
piece of a template and needs generalization.

The algorithm described finds those lines in a template that have targets of the
gates, but fails to extract all the possible assignments of the
controls. Generalization requires finding all the
possible gate controls that apply without changing the network
functionality, {\em i.e.} leaving it the identity. 
The following Theorem is useful 
as it limits the set of choices one can make to assign
the controls.  

\begin{definition}
For any network $G_0G_1...\;G_{m-1}$ with an input line that has
controls only (control line), its {\bf characteristic vector}
$(\alpha_0,\alpha_1,...\;\alpha_{m-1}),\; \alpha_i \in\{0,1\}$ for
$0 \leq i \leq m-1$ has ones at positions $i$ where the gate $G_i$
has a control, and zeros everywhere else.
\end{definition}

\begin{theorem}
\begin{enumerate}
\item If a control line with the characteristic
vector $(\alpha_0,\alpha_1,...\;\alpha_{m-1})$ appears in a
template of size $m$, any set of lines with this characteristic
vector is a valid control set.

\item Lines with characteristic vectors $(0,0,...\;0)$  and
$(1,1,...\;1)$ are valid control lines for any template.

\item If lines with characteristic vectors $(\alpha_0,\alpha_1,...\;\alpha_{m-1})$
and $(\beta_0,\beta_1,...\;\beta_{m-1})$ are control lines of a template,
the line with characteristic vector
$(\alpha_0\vee \beta_0,\alpha_1\vee \beta_1,...\;\alpha_{m-1}\vee \beta_{m-1})$
is also a valid control line.

\item If there exists a line with exactly two
EXOR symbols on it, being targets of two gates $G_i$ and $G_j$,
every valid control line has $\alpha_i = \alpha_j$.
\end{enumerate}
\end{theorem}

\begin{proof}

\noindent 1.\;\;\;\;\; To prove the statement we want to check if the operation of repeating
the number of controls of certain type keeps the identity being the identity.
Assuming $(\alpha_0,\alpha_1,...\;\alpha_{m-1})$ is a valid control line
labeled $x_1$ create line $x_2$ of the same type and show that it is a valid
control. There are two cases to prove. First, $x_2=0$ and second, $x_2=1$.
For $x_2=0$ the network is equivalent to the network without gates $G_i$ for
every $\alpha_i=1$. Same thing happens if line $x_1$ is set to zero. 
In case $x_2=1$ variable $x_2$ can
be completely ignored, which does not change the network functionality.
Note, that when any control line is deleted from the template, the resulting
cascade still realizes the identity function. This observation is useful in
understanding of how a large gate with many controls as shown in the
template in Figure \ref{tem7inv} can match a relatively
small gate in a network to be simplified: the control lines indicate which
controls are possible, but may not be necessary for a specific matching.

\noindent 2.\;\;\;\;\; Line with characteristic vector $(0,0,...\;0)$ is a ``virtual" line
whose presence or absence does not change anything. Thus, it can as well
be a control. Consider line $x$ with characteristic vector $(1,1,...\;1)$.
For $x=0$ all the gates do nothing as a control value of zero results in
zero value of the corresponding product, and no target line changes its
value. Case $x=1$ is equivalent to having a line with zero characteristic
vector.

\noindent 3.\;\;\;\;\; Assume line $x_1$ has characteristic vector
$(\alpha_0,\alpha_1,...\;\alpha_{m-1})$ and line $x_2$ has characteristic
vector $(\beta_0,\beta_1,...\;\beta_{m-1})$. Create line $x_3$ with the
characteristic vector
$(\alpha_0\vee \beta_0,\alpha_1\vee \beta_1,...\;\alpha_{m-1}\vee \beta_{m-1})$.
We want to prove that setting its values to zero and one does not change
the network functionality, that is, it stays the identity. Setting $x_3$
to $1$ is equivalent to ignoring value on this line, thus the network
will realize the identity. Setting $x_3$ to $0$ is equivalent to setting
both $x_1$ and $x_2$ to zero. Since $x_1$ and $x_2$ are valid controls,
setting them to zero does not change the network output (the network will
stay the identity). This means that setting $x_3$ to $0$ keeps the network
output being equal to its input, so the property of being the identity is
conserved, and $x_3$ is a valid control.

\noindent 4.\;\;\;\;\; Prove by contradiction. Suppose there is a control line $x$ with
characteristic vector $(\alpha_0,\alpha_1,...\;\alpha_{m-1})$ such
that $\alpha_i \neq \alpha_j$ and prove that in such case the network does
not realize the identity. Without loss of generality assume that
$\alpha_i=0$ and $\alpha_j=1$. Use Observation 4 to transform network
$G_0G_1...\;G_{m-1}$ to the form $G_iG_{i+1 \bmod m}...\;G_{i-1 \bmod m}$.
These two networks can be identities only simultaneously. Now, set
the input pattern to have ones on every bit except bit $x$, whose
value is set to $0$. Then, gate $G_i$ flips the value of its target
bit $y$ from $1$ to $0$. The only other gate that affects bit $y$
is $G_j$. Since controlling bit $x$ of the gate $G_j$ is zero (when
propagated control bits do not change their values, and $x$ has
controls only), gate $G_j$ will not flip its target value $y$. Thus,
at the end of the network bit $y$ will arrive flipped, which
contradicts the statement that network
$G_iG_{i+1 \bmod m}...\;G_{i-1 \bmod m}$ is the identity.
\end{proof}

\begin{figure}[tb]
\centering
\includegraphics[height=35mm]{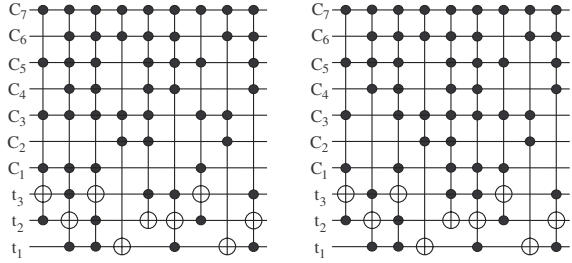}
\caption{Generalizations of a size 9 template.} \label{tem9}
\end{figure}

Template generalization is a part of our software package.
It is interesting to note that during the generalization process the
number of templates may increase. Figure \ref{tem9} illustrates how
a template found by our program (bottom 4 lines: $t_1,t_2,t_3$ and 
$C_1$, $|C_1|=1$) can be generalized in two different ways.

In \cite{ar:mdmCAD}, we reported a Toffoli network with 4 gates
for the 3-bit binary full adder. Assuming the trivial cost metric,
we applied our templates. This resulted in no gate count reduction 
and we can conclude that the network is optimal for the given cost metric (gate count).
Prove by contradiction. Suppose it is not. Then, there exists a smaller network for a 3-bit
full adder, say with 3 gates. Using Observation 2 one can construct
an identity cascade of size 7 built on 4 lines that would differ from the templates and will
not be simplified by the means of the templates. Running our template 
finding program shows that it is impossible, and hence
the network is optimal. The following theorem generalizes this observation.

\begin{theorem} \label{th:temm}
For the complete classification of the templates of size
up to $m$ and their complete (in the sense that no possible application 
is missed) application to network size reduction:
\begin{itemize}
\item For even numbers $m$, each sub-network
of size $\frac{m}{2}$ is optimal in any metric. The network itself
is optimal if the number of gates is $\frac{m}{2}$ or less.

\item For odd numbers $m$, each sub-network of
size $\lfloor \frac{m}{2} \rfloor$ is optimal in any metric, and
each sub-network of size $\lceil \frac{m}{2} \rceil$ is optimal in
the trivial metric. \noindent Similar statements hold for the
entire network if the number of gates is not greater than $\lfloor
\frac{m}{2} \rfloor$ or $\lceil \frac{m}{2} \rceil$ respectively.
\end{itemize}
\end{theorem}

\begin{figure}[tb]
\centering
\includegraphics[height=46mm]{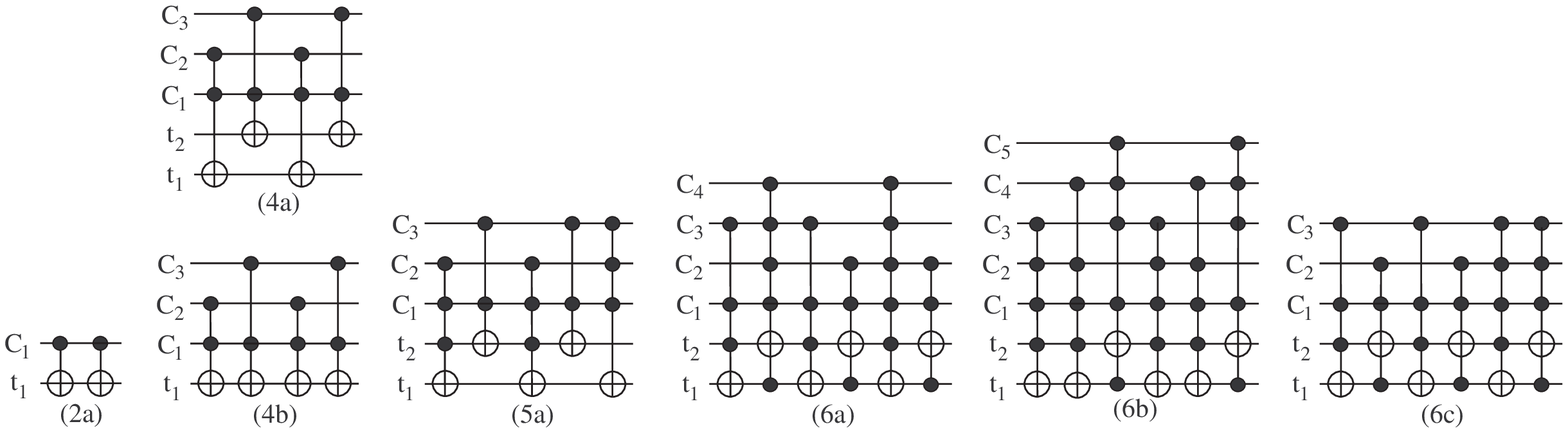}\\
\includegraphics[height=90mm]{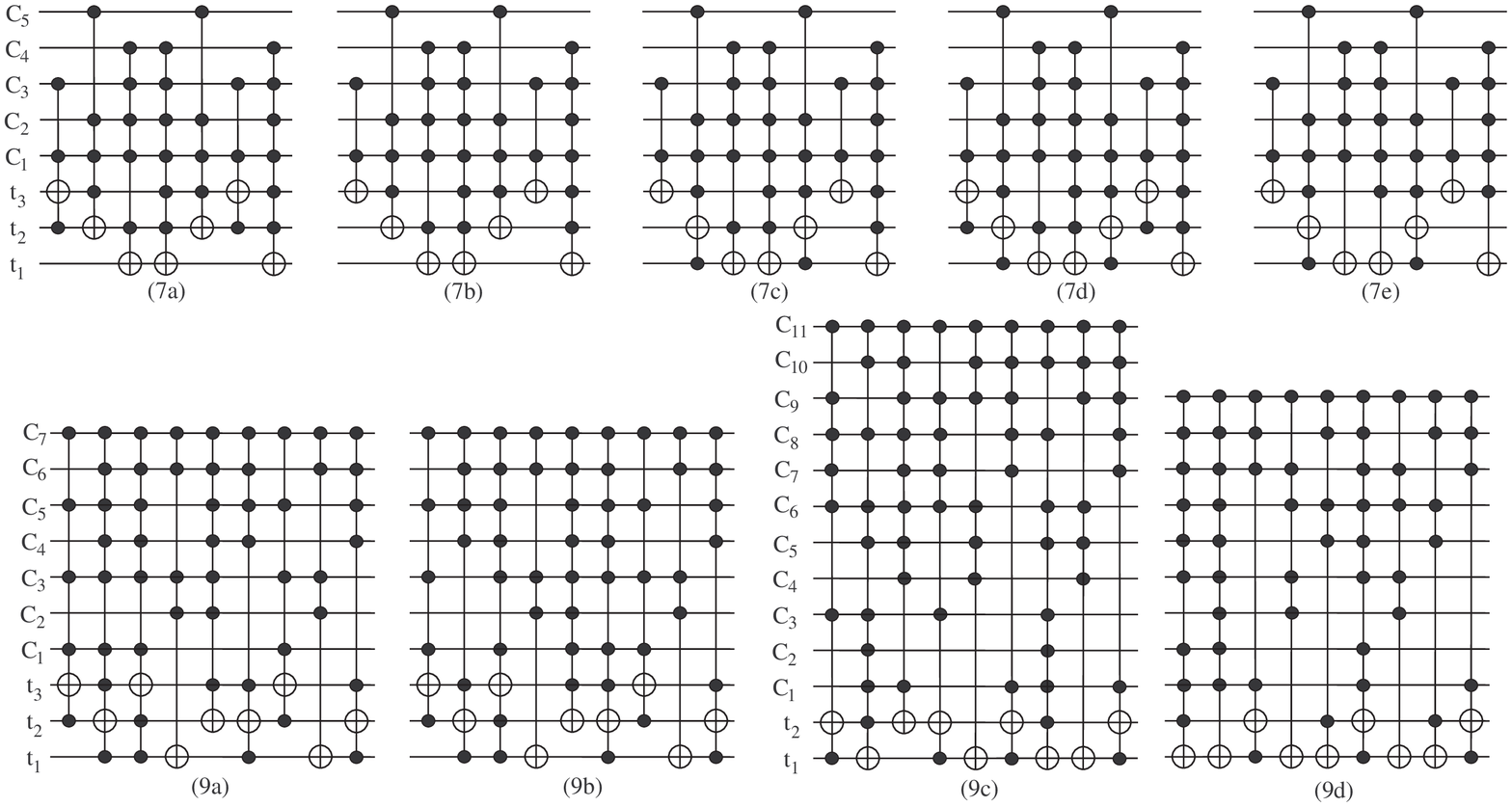}
\caption{Templates of size 7, and some templates of size 9.} \label{alltemplates}
\end{figure}

We conclude this subsection with a (most likely, complete) 
list of the templates of size up to 7  
and some templates of size 9, illustrated in Figure \ref{alltemplates}. Lines
$t_i$ in Figure \ref{alltemplates} represent each a single line, and lines 
marked with $C_j$ represent a possibly empty set of lines, all of the same form.
We note that the templates of size less than 5 are equivalent to those found in 
\cite{ar:mdmCAD}. We report a smaller number of different templates of size 6, 
as compared to the templates reported in \cite{ar:mdmCAD}.  
There are two reasons for that. First, the template illustrated in Fig. 8(d) of
\cite{ar:mdmCAD} is undergeneralized, which we found with the help of our new 
software. And second, the template classification depends on how the templates 
are applied.
Our algorithm for template application differs from the originally reported 
(\cite{ar:mdmCAD}) and is discussed in the following subsection. A quick explanation 
of why the new algorithm is more accurate at finding more simplification 
than the original, is that template 
illustrated in Fig. 8(b) of \cite{ar:mdmCAD} can 
now be simplified with the other templates, which 
was not possible before. This is due to the improved matching algorithm.

\subsection{Template Application}\label{sec:tem}
To apply templates to circuit cost reduction, we first consider
all the templates of the form $ABAB$. Such templates applied for parameter $p=2$
result in the rewriting rule $AB \rightarrow BA$. That is,
they define when the two adjacent gates in a cascade can be swapped. We call
such templates {\bf moving rules} and apply them to move the gates in
a cascade to permit the application of cost reducing template substitutions. It 
transpires that for the most network types considered in the literature (binary
reversible, MVL reversible, and quantum) the complete description
of the templates of form $ABAB$ is very simple (this may of course not be true 
for all gate types). Assuming gate $A$ has control(s) $C_A$ and target $T_A$
and gate $B$ has control(s) $C_B$ and target $T_B$ these two gates
form a moving rule if, and only if, $T_A \not\in C_B$ and $T_B\not\in C_A$.

Templates of other size than 4 (and all Toffoli templates of size 4 are the moving rule) 
are applied as follows. We choose a starting gate for matching. 
The position of the starting gate ($Start$) in the 
matching will change with time, and we begin with $Start=2$. 
Suppose $Start=k$ in a cascade with $n$ gates 
at the present time. We apply smaller templates first. They are easier to match,
because one needs to find less gates to do the replacement, and in a sense smaller templates 
allow more general network transformation (for instance, applying size 2 templates 
can be thought of as deleting pairs of equal gates, while applying size 9 templates 
is hard to describe by words). For each of the templates, we match gate $k$ in the 
network to the first gate in each of the $m$ cycles of the template, which is always possible. 
We then try to find the gates in the network that match those in the template 
assuming the first gate of the template cycle matches this $k^{th}$ gate in the network
and trying both directions for the template application. Next we only explain how to apply 
a single template cycle in forward direction, because application of other cycles
and in backward direction is analogous. At this point, we create two 
arrays, integer $MatchIndex[\;]$ with one element $k$ indicating that one gate at position
$k$ in the network is found and properly matched, 
and Boolean $MoveIndex[\;]$ with one element equal 1 and indicating that all gates can be 
moved to the one found (in this case no moving is required). In addition, integer 
$CurrentGate=k$ indicates that at the present moment we look at the gate $k$.
To match more gates, we decrease $CurrentGate$ by 1 and see if gate 
$k-1$ in the network matches the second gate in the template cycle. If it does,
we increase the size of $MatchIndex$ array by 1, and add $k-1$ to it. 
We increase size of the $MoveIndex$ array and add a new element, 1 to it. 
Since gate $k-1$ neighbors with gate $k$, there is no need to check if the gates 
can be moved together. Finally, we check if these 2 gates can be replaced with a smaller 
network using the present template cycle, 
and if they can, do the replacement and return $Start=k-1$.
The template matching resumes with starting gate at position $k-1$ and by trying 
the smallest template first. If gate $CurrentGate$ did not match the second gate in the template,
we decrease integer $CurrentGate$ by 1 (it is now equal $k-2$) and see if this gate matches
the second gate of the template cycle. 

In general, if some $s$ gates are matched and can be moved together
(that is, $MatchIndex=[k_1,k_2,...,k_s]$ and $MoveIndex=[m_1,m_2,...,m_s]$ 
where $MoveIndex$ contains a non-zero value indicating that the gates can be moved 
to the corresponding position), and a gate in a network at position $CurrentGate=k_{s+1}$ 
matches $(s+1)^{st}$ gate of the template cycle, the procedure for matching is as follows. 
First, we check if the gates can be moved together to each of the network positions 
$k_1,k_2,...,k_s,k_{s+1}$. If the gates can be moved together, we create array 
$MoveIndex$ with $s+1$ Boolean values showing where it is possible to move the 
gates. We next check if it is beneficial to replace the matched 
and movable together $s+1$ gates with the remaining $m-s-1$ of the given template. 
If it is, we do the replacement at the maximal value of an element of $MatchIndex$, $k_j$,
corresponding to the non-zero value $m_j$ of the $MoveIndex$ and return new $start=k_j-s+j$.
The template matching resumes from this position in the network and the smallest template.
If the gates cannot be moved together, we decrease value of the $CurrentGate$ by one 
and try to match gate at this position in the network. When $CurrentGate$ becomes equal zero 
or if we cannot match enough gates to do a beneficial replacement 
using the template, we try to match another cycle,
next template of the same size, or a larger template. If no templates match with 
a starting position $Start$, we increase its value by one (start matching with the
next gate in the network) until we run out of gates in the network that could serve
as a $Start$. In such case, template application is completed.

We illustrate how the templates are applied with an example below.

\begin{example}

\begin{figure}[tb]
\centering
\includegraphics[height=30mm]{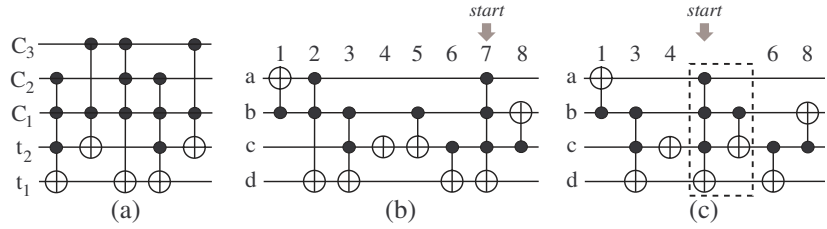}
\caption{Application of template (a) to the network (b) starting at 
the gate 7. The network simplifies to the form (c).} \label{temapply}
\end{figure}

Consider network in Figure \ref{temapply}(b). Suppose $Start=7$ and the template 
cycle that we want to match and apply is as illustrated in Figure \ref{temapply}(a). 
In the beginning of the matching we have $MatchIndex=[7],\;MoveIndex=[1]$ and
$CurrentGate=7$. Line $t_1$ of the template must correspond to the line $d$ of the 
network, and line $t_2$ should match one of $a$, $b$ or $c$ --- this guarantees that 
gate 7 matches the first gate of the template cycle.
The steps of matching are:
\begin{itemize}
\item Let $CurrentGate=7-1$. Gate $6$ does not match the second gate 
of the template cycle in Figure \ref{temapply}(a) since we expect to 
find a gate with target at a line where gate 7 has a control. Nothing is 
done, $CurrentGate$ is decreased by 1.
\item $CurrentGate=5$. Gate 5 matches the second gate of the template if 
$t_2=c$, $C_1=\{b\}$ and $C_3=\emptyset$. Gate 7 can be moved to gate 5, 
and gate 5 cannot move to the gate 7. Therefore, $MoveIndex$ becomes $[1,0]$.
The $MatchIndex$ is $[7,5]$. We check that the replacement of the two gates we matched
with the three reconstructed from the template is not beneficial, but since 
$MoveIndex$ has non-zero values, we try to match more gates.
\item $CurrentGate=4$. Gate $4$ in the network does not match the third gate in the 
template cycle because we are looking for a gate with the target on line $d$.
\item $CurrentGate=3$. Gate $3$ in the network does not match third gate in the 
given template cycle because we try to find a gate with no control at line $c$.
\item $CurrentGate=2$. Gate 2 matches the third gate of the template cycle if 
$C_2=\{a\}$. Gate 2 can be moved to gate 5, but gate 5 cannot be moved past gate 3.
Thus, $MoveIndex=[0,1,0]$. $MatchIndex=[7,5,2]$ and according to the template cycle,
these three gates can be replaced with two. It is clearly beneficial to do the replacement.
According to the $MoveIndex$ the replacement can be done if all gates are moved to the gate
5 in the network. The network after template 
application is illustrated in Figure \ref{temapply}(c). Return $Start=4$, because at this 
position the replacement part (in a dashed box) starts.
\end{itemize}
The template application resumes starting with the forth gate in the network 
in Figure \ref{temapply}(c) and trying to apply the smallest template.
\end{example}

In our program realization, function $apply\_templates$ is used to apply templates.
It has an option of applying the templates to reduce the gate count 
or the quantum cost and works according to the algorithm discussed above. 
We made a modification of the matching algorithm in which we never look for 
the gates in the network further away from position $start$ than 20. This 
is because we found that in practice the gate span in template application 
is usually less than 20. Such restriction also makes the template algorithm
faster --- it is linear in the number of gates in the network. The template 
application algorithm from \cite{ar:mdmCAD} has an $n^3$ worst 
case scenario and $n^2$ best case scenario runtime in terms of the gate count 
of the circuit to be reduced. The new template application algorithm introduced
in this section reduces the circuits better (assuming both algorithms work 
with the same set of templates) than the one presented in \cite{ar:mdmCAD},
and can be used in conjunction with different circuit cost metrics.

%

\section{Resynthesis Procedure}\label{sec:resynth}

In our program implementation, we first
synthesize a function and its inverse using the MMD method \cite{ar:mdmCAD} and the
newly presented Reed-Muller spectra based algorithms. We then
simplify each of the synthesized networks using the templates,
choose the smaller network $N$ and declare it to be the final
implementation. Each subnetwork $N_{sub}$ of the final
implementation is itself a network computing some reversible
function. This reversible function can be determined and
synthesized on its own. If such resynthesis yields a smaller
subnetwork, it replaces $N_{sub}$ leading to simplification of the
overall network $N$.

We have implemented two drivers for this resynthesis procedure.
First, a $random\_driver$ which performs a user-specified number of
iterations. For each iteration, a number (again specified
by the user) of random subnetworks are resynthesized and the best overall
simplification is chosen and forwarded to the next iteration.
Second is an $exhaustive\_driver$. It tries all possible subnetworks
with at least 5 gates of a given network. The requirement of 5
gates is because it is not necessary to resynthesize networks with
4 or less gates, since in our synthesis approach every subnetwork
of length 4 is optimal. This result is a corollary of Theorem \ref{th:temm}.

When we synthesize networks, random driver is used first. When it 
does not simplify the network after a few iterations, we run the exhaustive
driver (time allowing) to make sure that no sub-network simplifies. The exhaustive
driver can take a long time, especially if applied to larger complex functions
such as $hwb11$. We did not apply exhaustive driver to the functions of size 
16 and greater. Note that using a random driver results in
different scenarios for network simplification and the simplified
network may differ from one application to the next. It is
expected that some of the larger networks considered in Section
\ref{sec:r} may be further reduced by multiple applications
of the random driver.

\section{Results}\label{sec:r}

In the literature, one of the common tests of the quality of a
reversible synthesis method is how it performs on the 40,320
$3\times 3$ reversible functions \cite{ar:mdmCAD, co:aj, co:k,
ar:mdmCAD}. We used the 3 synthesis methods that are applied to both
function and its inverse, then the templates were 
applied and the $exhaustive\_driver$ is run 
until no further simplification is found. This is a time consuming test, 
and it takes around 96 hours for it to complete. Techniques to reduce 
the runtime are discussed in Section \ref{sec:fw}.

Table \ref{tab-sizes} compares our synthesis results
to the earlier reported synthesis algorithms and the optimal
results found by depth-first search. It can be seen that our
results are significantly closer to the optimal synthesis than the
basic MMD algorithm plus templates of maximal size 6
\cite{ar:mdmCAD} (column {\bf MMD}), and over twenty times (overhead) as
close to the optimum as a recently presented Reed-Muller based
tree search algorithm \cite{co:aj} (column {\bf AJ}). Our results
are, on average (WA), only $0.16\%$ off from the optimal size (column
{\bf Opt.} \cite{ar:spmh}). It can also be seen that our synthesis
results are better than the best presented by Kerntopf \cite{co:k}
(column {\bf K}), even though that his work uses a larger gate
library (given a large gate library one would expect lower gate
counts).

\begin{table}
\begin{center}
\begin{tabular}{|r|r|r|r|r||r|}\hline
{\bf Size} & {\bf MMD} &{\bf AJ}&{\bf Ours}&{\bf Opt.}& {\bf K} \\ \hline
13   &  6      &       &       &       &    \\
12   &  62     &       &       &       &    \\
11   &  391    &       &       &       &    \\
10   &  1444   &       &       &       &    \\
9    &  3837   & 30    & 2     &       & 86   \\
8    &  7274   & 3297  & 659   & 577   & 2740   \\
7    &  9965   & 12488 & 10367 & 10253 & 11774  \\
6    &  9086   & 13620 & 16953 & 17049 & 13683  \\
5    &  5448   & 7503  & 8819  & 8921  & 8068   \\
4    &  2125   & 2642  & 2780  & 2780  & 3038   \\
3    &  567    & 625   & 625   & 625   & 781    \\
2    &  102    & 102   & 102   & 102   & 134    \\
1    &  12     & 12    & 12    & 12    & 15 \\
0    &  1      & 1     & 1     & 1     & 1  \\ \hline\hline
WA:  & 6.801   & 6.101 & 5.875 & 5.866 & 6.010  \\  \hline
\%   & 116\%   & 104\% & 100.16\% & 100\% & N/A    \\ \hline
\end{tabular}
\end{center}
\caption{Number of reversible functions using a specified number
of gates for $n=3$. Column {\bf K} is separated from the remainder of the table 
because the gate library used in that work is different (larger).}
\label{tab-sizes}
\end{table}

Running a synthesis algorithm on all size 3 reversible functions
could be an interesting test, but it does not illustrate how the
synthesis method applies to large specifications, whose synthesis
is the main reason to design an automated procedure. We have
applied our synthesis approach to a number of reversible benchmark
specifications from \cite{wp}\footnote{
In our comparison, we considered the networks and function specifications 
from the above web page. However, our quantum cost calculation differs 
from the one used in \cite{wp}, therefore quantum costs reported in 
Table \ref{bs} are slightly different from those that can be found online.}. 
We report two types of the results:
we minimize the gate count and we minimize the quantum cost separately.
The results are given in Table \ref{bs}. The {\bf name}, 
{\bf size}, {\bf GC} and {\bf QC} columns give the name of
each benchmark function, its size (number of variables) of the
reversible specification as considered in the literature, the 
best reported gate count, and the best reported quantum cost for 
the networks with Toffoli gates. 
Next two columns report the gate count and the quantum cost when 
our tool is applied to synthesize a given function with the option of 
minimizing the gate count. The last two columns report the synthesis 
results with the option of quantum cost minimization. We find that 
realizations in the last two columns could be more practical. We note
that networks for benchmark functions $4mod5$, $5mod5$, $hwb8-hwb11$, and 
network for $rd53$ with quantum cost 79 found in \cite{wp} are 
the results of the techniques discussed in this paper and were not 
reported before. 

Table \ref{bs}
shows our software synthesize smaller\footnote{ \cite{wp} contains networks synthesized
using Toffoli and Fredkin gates, but we do not compare our results to 
those in a table form, just mention that the newly presented results are, generally, 
significantly better.} networks than earlier
presented heuristics. For instance, the gate count for the $hwb6$
benchmark function was reduced from 126 to 42 gates, that is, our
network is one third of the size of the best previously
presented; and quantum cost for an implementation of this 
function was reduced more than 10 times. 

We limited the search time for our software to 12 hours for each benchmark 
function. Most functions took significantly less time to synthesize than the 
allowed 12 hours; most time (12 h) was spend to synthesize only one function,
$cycle18\_3$. A general rule was to synthesize a function using all three algorithms,
apply the templates, resynthesize with $random\_driver$ until several iterations 
do not bring any simplification and apply $exhaustive\_driver$ until no 
further simplification. In the chosen period of 12 hours, there was no time left to 
apply $exhaustive\_driver$ to functions (networks for) $hwb7$, $hwb8$, $hwb9$ and all 
networks with 10 and more variables other than $ham15$ and $cycle10\_2$. 
Due to the time constraints, we 
did not apply $random\_driver$ to the networks for $hwb11$ and 
$cycle18\_3$. Our software potentially can synthesize functions with more than 
21 variables, but as the number of variables and gates in the synthesized 
network grows, the runtime for such synthesis grows exponentially.

\begin{table}
\begin{center}
\begin{tabular}{|c|c||c|c||c|c|c|c|} \hline
\multicolumn{4}{|c||}{{\bf Benchmark/its best circuit:}} &
\multicolumn{2}{|c|}{{\bf Gate count minimization:}} &
\multicolumn{2}{|c|}{{\bf Quantum cost minimization:}} \\ 
{\bf name} & {\bf size}& {\bf GC} & {\bf QC} & {\bf GC-gc} 
& {\bf QC-gc} & {\bf GC-qc} & {\bf QC-qc} \\ \hline
3\_17        & 3  & 6   & 14      & 6   & 14   & 6   & 14 \\ \hline
4\_49        & 4  & 16  & 64*     & 12  & 32   & 12  & 32 \\ \hline
4mod5        & 5  & 5   & 13      & 5   & 13   & 5   & 13 \\ \hline
5mod5        & 6  & 10  & 85*     & 8   & 77* & 10 & 71* \\ \hline
add3         & 4  & 4   & 12      & 4   & 12  & 4  & 12 \\ \hline
cycle10\_2   & 12 & 19  & 1206  & 19 & 1206 & 19 & 1206 \\ \hline
cycle17\_3   & 20 & 48  & 6069  & 48 & 6069 & 48 & 6069 \\ \hline
cycle18\_3   & 21 & N/A & N/A   & 51 & 6819 & 51 & 6819 \\ \hline
ham3         & 3  & 5   & 9       & 5   & 9   & 5  & 9 \\ \hline
ham7         & 7  & 23  & 91      & 21  & 69  & 25 & 49 \\ \hline
ham15        & 15 & 132 & 1881    & 70  & 463 & 109 & 214 \\ \hline
hwb4         & 4  & 17  & 69*     & 11  & 23  & 11 & 23 \\\hline
hwb5         & 5  & 55  & 353*    & 24  & 114 & 24 & 114\\ \hline
hwb6         & 6  & 126 & 1519*   & 42  & 150 & 42 & 150 \\ \hline
hwb7         & 7  & 289 & 5196*   & 236 & 3984* & 331 & 2609* \\ \hline
hwb8         & 8  & 637 & 14636*  & 614 & 12745* & 749 & 6197* \\ \hline
hwb9         & 9  & 1544 & 43138* & 1541 & 43089* & 1959 & 20378* \\ \hline
hwb10        & 10 & 3631 & 120034* & 3595 & 117460* & 4540 & 46597* \\ \hline
hwb11        & 11 & 9314 & 328200* & 8214 & 336369* & 11600 & 122144*\\ \hline
mod5adder    & 6  & 21  & 145   & 15 & 91 & 17 & 81\\ \hline
mod1024adder & 20 & 55  & 1575  & 55 & 1575 & 55 & 1575\\ \hline
rd53         & 7  & 12  & 128   & 12 & 128 & 16 & 67 \\ \hline
rd53         & 7  & 16  & 79    & 12 & 128 & 16 & 67 \\ \hline
%
\end{tabular}
\caption{Benchmark function synthesis. Actual circuits are available from \cite{wp}.}
\label{bs}
\end{center}
\end{table}

\section{Future Work}\label{sec:fw}

Our program realization of the discussed methods is no way 
optimized. We use a resource demanding truth table representation 
of a function, which, in addition to slowing the software 
significantly limits the scalability of our implementation. To date,
we found that scalability of our approach is satisfactory, however,
a more robust function representation must be employed in the future 
to minimize circuits for the functions with more than 21 variables.

Further work has to be done to optimize the code. For instance, 
our algorithm can be easily parallelized. Assuming one has a 
6 processor machine, each of the 6 networks (3 methods, function 
and its inverse are synthesized) can be synthesized (including the 
template application) on a separate processor. Work of 
$random\_driver$ and $exhaustive\_driver$ can be distributed evenly
among the processors. In total, such algorithm on a parallel machine
should be able to run almost 6 times faster as compared to a 
single processor machine. For large networks,
template application can be parallelized by cutting them into small 
sub-networks and then applying the templates at the cutting points by
restricting $Start$ to grow no more than 20.

Synthesis of incomplete specifications (Boolean multi output functions) 
is possible using each of the the newly presented synthesis methods, 
as well as with the old one. This has to be investigated further since
most of the real world benchmark functions are irreversible and 
transforming function specification into a reversible before synthesizing it
should not be more effective than the straight synthesis of the multi output 
specification.

\section{Conclusion}
In this paper, we presented novel techniques for the synthesis of
reversible Toffoli networks.  The main contributions include two
Reed-Muller spectra based approaches to reversible synthesis; a
better characterization of templates and an improved method of 
their application, classification of the templates of size 7 (most likely, 
complete) and demonstration of some useful templates of size 9.
We also investigated a new approach involving resynthesis of
subnetworks that significantly improves the results, particularly
for larger benchmark functions. We structured our software as to 
have an option of minimizing the gate count or a technology-motivated
cost. To our knowledge, this is the first attempt to minimize 
technology motivated cost of the implementation in the relevant literature.

We have implemented our methods in C++ and shown they produce
results significantly better than those reported in the literature.  
Finally, we have identified several ways to improve this work.

\bibliographystyle{abbrv}

\end{document}